\documentclass[twocolumn]{aastex7}

\shorttitle{Simulation of the Solar Surface Magnetic Field from 2010 to 2024}
\shortauthors{Wang, Jiang, \&Luo}

\graphicspath{{./}{figures/}}

\usepackage{CJK}
\usepackage{multirow}
\usepackage{booktabs}
\usepackage{amsmath}
\usepackage{threeparttable}
\usepackage{verbatim}

\usepackage{hyperref}

\begin{document}
\begin{CJK*}{UTF8}{gbsn}

\title{Solar Surface Magnetic Field Simulation from 2010 to 2024 and Anomalous Southern Poleward Flux Transport in Cycle 24}

\correspondingauthor{Jie Jiang}
\email{jiejiang@buaa.edu.cn}

\author[orcid=0000-0003-0256-8102,sname=Ruihui,gname=Wang]{Ruihui Wang (王瑞慧)}
\affiliation{School of Space and Earth Sciences, Beihang University, Beijing, People’s Republic of China}
\affiliation{Key Laboratory of Space Environment Monitoring and Information Processing of MIIT, Beijing, People’s Republic of China}
\email[show]{wangruihui@buaa.edu.cn}  

\author[orcid=0000-0001-5002-0577,sname=Jie,gname=Jiang]{Jie Jiang (姜杰)}
\affiliation{School of Space and Earth Sciences, Beihang University, Beijing, People’s Republic of China}
\affiliation{Key Laboratory of Space Environment Monitoring and Information Processing of MIIT, Beijing, People’s Republic of China}
\email[show]{jiejiang@buaa.edu.cn}  

\author[orcid=0000-0002-6977-1239,sname=Yukun,gname=Luo]{Yukun Luo (罗昱琨)}
\affiliation{School of Space and Earth Sciences, Beihang University, Beijing, People’s Republic of China}
\affiliation{Key Laboratory of Space Environment Monitoring and Information Processing of MIIT, Beijing, People’s Republic of China}
\email{luoyukun@buaa.edu.cn}

\begin{abstract}
The solar surface magnetic field is fundamental for modeling the coronal magnetic field, studying the solar dynamo, and predicting solar cycle strength. We perform a continuous simulation of the surface magnetic field from 2010 to 2024, covering solar cycle 24 and the ongoing cycle 25, using the surface flux transport model with assimilated observed active regions (ARs) as the source. The simulation reproduces the evolution of the axial dipole strength, polar field reversal timing, and magnetic butterfly diagram in good agreement with SDO/HMI observations. Notably, these results are achieved without incorporating radial diffusion or cyclic variations in meridional flow speed, suggesting their limited impact. Poleward surges of the following polarity typically dominate throughout the cycle, but in the southern hemisphere during cycle 24, they are limited to a short period from 2011 to 2016. This anomalous pattern arises from intermittent AR emergence, with about 46\% of total unsigned flux contributed by ARs emerging during Carrington Rotations 2141-2160 (September 2013 - February 2015). These ARs show a strong active longitude at Carrington longitudes $200^\circ$-$260^\circ$ and a weaker one at $80^\circ$-$100^\circ$. After 2016, poleward migrations of leading-polarity flux become dominant, despite most ARs following Joy's and Hale's laws. This reversal is likely due to prolonged intervals between AR emergences, which allow leading-polarity flux to distribute across a broad latitude range before cancellation by subsequent ARs. These findings highlight the importance of the temporal interval of AR emergence in driving the flux transport pattern. 

\end{abstract}

\keywords{\uat{Solar physics}{1476}---\uat{Solar cycle}{1487}---\uat{Solar magnetic fields}{1503}---\uat{Solar surface}{1527}--- \uat{Solar active regions}{1974}---\uat{Astronomy databases}{83}}

\section{Introduction} \label{sec:intro}

The solar surface magnetic field is an important and relatively well-observed component of the solar magnetic field. It provides the inner boundary condition for modeling the coronal magnetic field \citep{Yeates2007, Yeates2023}, and the outer boundary condition for solar dynamo models. The polar field at solar minimum has also been shown to be a good predictor of the strength of the following solar cycle \citep{Jiang2007, Bhowmik2018, Petrovay2020SCpd}.

Solar active regions (ARs) are the primary source of the surface magnetic field. After emergence, AR flux is transported by surface flows, including differential rotation, meridional flow, and supergranular diffusion. Ultimately, some of the AR flux is carried to the poles, contributing to the evolution of the polar field \citep{Leighton1964, Wang1991}. This flux transport process is usually described by the surface flux transport (SFT) model \citep{WangYM1989ApJ, Jiang2014ssr, Yeates2023}. The SFT model has successfully reproduced various observed features of the solar surface magnetic field, including the magnetic butterfly diagram, the poleward transport of AR flux (surges or plumes), the solar axial dipole strength (dipole moment), and the polar field \citep{Jiang2014apj, Jiang2015, Lemerle2015, 2020Yeates}.

However, when using the SFT model to simulate multiple cycles, \cite{Schrijver2002} suggested that an additional decay term is necessary to maintain consistency between simulated and observed polar fields. This decay term represents the radial diffusion of magnetic flux \citep{Yeates2023}. Some simulations have instead adjusted the meridional flow speed to match the simulated polar field or solar axial dipole strength with observations \citep{WangYM2002, WangYM2009}. However, the cycle-dependent variations in meridional flow speed required by these simulations are not yet supported by observations.

Unlike adding a radial diffusion term or adjusting transport parameters, some simulations achieve a good match for the polar field by modifying the source parameters. Most past SFT simulations use symmetric bipolar magnetic regions (BMRs) as the source. Observations show that BMR parameters vary with cycle strength. \cite{Dasi-Espuig2010} reports the variation in sunspot group tilt angle with cycle strength, which has been further confirmed by subsequent studies \citep{Jiang2011, Jiao2021}. Incorporating the observed tilt variation, \cite{Cameron2010} successfully simulated the surface magnetic field and open flux for cycles 15-21. Additionally, average sunspot emergence latitudes also vary with cycle strength \citep{Waldmeier1955, Solanki2008}. \cite{Jiang2014apj} shows that ARs emerging closer to the equator have a stronger impact on the polar field. Therefore, the cyclic variation in emergence latitude can also influence polar field strength \citep{Jiang2020, Talafha2022}.

However, BMRs are insufficient to accurately represent ARs. The evolution of ARs with complex configurations often differs significantly from their simplified BMR representations \citep{Jiang2019, Yeates2023}. As a result, BMR approximations can lead to inaccurate estimates of ARs’ impact on the polar field, often overestimating the total contribution of all ARs within a solar cycle \citep{2020Yeates, WangZF2021, Wang2024}. A more accurate approach is to directly assimilate observed ARs, as adopted in many simulations \citep{Yeates2007, WangZF2020, YangSH2024}. Using ARs constructed from Ca II K synoptic maps, \cite{Yeates2025} successfully reproduced polar fields consistent with relevant proxy observations from 1923 to 1985. Notably, these simulations did not include radial diffusion or cycle-dependent variations in meridional flow speed, suggesting that source variation alone may be sufficient to account for the cyclic evolution of the surface magnetic field.

To enable the direct assimilation of observed ARs, a database of ARs detected from magnetograms is essential. Such a database is critical for accurately simulating and understanding the evolution of the surface magnetic field, as well as for predicting the polar field at solar minimum, which in turn forecasts the strength of the next solar cycle \citep{Petrovay2020SCpd}. To meet this need, we developed the Active Region database for Influence on Solar cycle Evolution (ARISE) in \cite{Wang2023} and \cite{Wang2024} based on synoptic magnetograms from the Michelson Doppler Imager on board the Solar and Heliospheric Observatory (SOHO/MDI; \citealt{MDI}) and the Helioseismic and Magnetic Imager on board the Solar Dynamics Observatory (SDO/HMI; \citealt{HMI}), covering the period from 1996 to 2024 across multiple solar cycles. ARs are identified using morphological operations and region-growing techniques to ensure robust detection. To exclude regions that do not represent new flux emergence, we remove over-decayed ARs, unipolar regions, and repeated detections of long-lived ARs. With appropriate detection and filtering, this database serves as a reliable, high-quality source for surface magnetic field simulations. 

AR flux migrations to the poles are episodic rather than smooth, with strong flows, known as surges or plumes, clearly visible in magnetic butterfly diagrams \citep{Howard1981}. Since following polarities generally emerge at higher latitudes, surges are typically dominated by them. However, in the southern hemisphere during cycle 24, HMI synoptic magnetograms show that following polarity surges were confined to a short period, with subsequent poleward migrations dominated by leading polarities, particularly during the declining phase. Although many studies have investigated cycle 24 surges \citep{Yeates2015, Sun2015, WangZF2020, Mordvinov2022}, the anomalous poleward migration behavior in the southern hemisphere remains understudied. Our ARISE database, covering nearly the full cycle 24 and including axial dipole strengths that quantify each AR’s contribution to the polar field, offers a valuable resource for analyzing this anomalous flux transport behavior.

In this paper, we continuously simulate the solar surface magnetic field from 2010 to 2024 by assimilating ARs in our database, without including radial diffusion or cycle-dependent variations in meridional flow speed. We further investigate the anomalous poleward flux transport in the southern hemisphere during cycle 24. The paper is structured as follows. Section \ref{sec:model} describes the SFT model and source assimilation method. Section \ref{sec:sml surfaceflux} presents the simulation of the surface magnetic field from 2010 to 2014 and its comparison with HMI observations. Section \ref{sec: c24 south} analyzes the anomalous southern poleward flux transport during cycle 24. Section \ref{sec:conclusion} summarizes and discusses the above results.

\section{surface flux transport model} \label{sec:model}

\subsection{Model Description}

The equation of the surface flux transport (SFT) model is 
\begin{equation}
\label{eq:SFT}
\begin{split}
\frac{\partial B}{\partial t} &= -\omega(\theta) \frac{\partial B}{\partial \phi} - \frac{1}{R_\odot \sin \theta} \frac{\partial}{\partial \theta} [u(\theta) B \sin \theta] \\
&+ \frac{\eta}{R_\odot^2} \left[ \frac{1}{\sin \theta} \frac{\partial}{\partial \theta} \left( \sin \theta \frac{\partial B}{\partial \theta} \right) + \frac{1}{\sin^2 \theta} \frac{\partial^2 B}{\partial \phi^2} \right]\\
&+ S(\theta, \phi, t),
\end{split}
\end{equation}
where B is the radial magnetic flux density, $\theta$ and $\phi$ are colatitude and heliographic longitude, respectively. The parameter $\eta$ denotes the supergranular diffusivity. The differential rotation $\omega(\theta)$ is taken from \cite{Snodgrass1983}. The meridional flow $u(\theta)$ is not well constrained by observations. Several commonly used profiles include those proposed by \cite{van_Ballegooijen1998}, \cite{Lemerle2017}, \cite{Whitbread2018}, and \cite{WangYM2017}. \cite{Petrovay2020AM} showed that meridional flows with the same equatorial divergence and peak speeds located away from low latitudes tend to yield comparable axial dipole strengths. For simplicity, we adopt the profile of \citet{van_Ballegooijen1998} in this study. The impact of different meridional flow profiles on the simulation results will be explored in future work. The exact formula of meridional flow is :
\begin{equation}
u(\lambda) = 
\begin{cases}
u_0 \sin \left( \pi \lambda / \lambda_0 \right) & \text{if } |\lambda| < \lambda_0 \\
0 & \text{otherwise},
\end{cases}
\end{equation}
where $\lambda$ is latitude, $\lambda_0 = 75 ^{\circ}$, and $u_0$ is the peak meridional flow speed. After trials, we adopt a supergranular diffusivity of $\eta = 340$ km$^2$/s and a meridional flow speed of $u_0 = 13$ m/s in this study. The combined effect of supergranular diffusion and meridional flow on the polar field is characterized by the dynamo effectivity range, defined as $\lambda _R =\sqrt{\frac{\eta }{R_{\odot}^2\Delta u}}$, where $\Delta u$ represents the divergence of the meridional flow at the equator, given by $\left. \frac{1}{R_\odot} \frac{du}{d\lambda} \right|_{\lambda = 0}$ \citep{Petrovay2020AM}. With the adopted parameters, we obtain $\lambda_R = 7.17^\circ$. The source term $S(\theta, \phi, t)$, representing surface magnetic flux emergence, is described in the next subsection.

To better match simulated polar fields with observations, some studies introduce a radial diffusion term $B/\tau$ into the SFT equation \citep{Schrijver2002, Whitbread2017}. However, we will show that this term is not essential and simulations can still reproduce observations accurately if the source term is properly assimilated.

We have developed a new code to solve the SFT equation based on spectral methods. The simulation is conducted at a spatial resolution of $180 \times360$ and a temporal resolution of 1 day. Further details on the numerical implementation are available in Luo et al. (2025, in preparation).

\subsection{Active Region Database and Source Assimilation}

The source term in our simulation is derived by assimilating ARs from the Active Region database for Influence on Solar cycle Evolution (ARISE; \citealt{Wang2023, Wang2024}). The database and associated codes are publicly available on GitHub \footnote{\texttt{AR database:} \url{https://github.com/Wang-Ruihui/A-live-homogeneous-database-of-solar-active-regions}.}, and version 3.0 is archived in Zenodo \citep{Database_WangRH_2025}. ARs are identified from SOHO/MDI and SDO/HMI synoptic magnetograms \citep{MDI, HMI}, covering Carrington Rotations (CRs) 1909-2290, corresponding to May 1996 through November 2024 and spanning solar cycles 23, 24, and part of 25.

Since some ARs can persist across multiple CRs \citep{Harvey1993} and usually significantly impact the polar field in solar minimum, we introduce a repeat-AR-removal module in \cite{Wang2024}. This module derotates detected ARs to the previous CR to correct for differential rotation, then compares the AR flux with the flux within the corresponding region in the earlier CR map. If the flux ratio (previous CR to current CR) exceeds a threshold $T_{rt1}$ and a specified polarity condition is met (see \cite{Wang2024}), the AR is identified as a repeat of an existing one and excluded. The polarity condition prevents ARs with opposite dominant polarity to those in the previous CR map from being misidentified as repeats.

However, the above procedure only removes repeated observations of ARs during their decay phase. Occasionally, ARs that are still in the process of emerging are captured in synoptic magnetograms. These non-fully-emerged ARs should also be excluded, as they are likely to reappear on subsequent maps with increased unsigned flux. To address this, we apply a similar approach to the repeat-AR-removal module. Specifically, each AR is rotated forward to the next CR, and the ratio of the flux within the corresponding region in the next CR map to the current AR flux is calculated. ARs are excluded if this ratio exceeds the threshold $T_{r2}$ and a polarity condition similar to that used in repeat-AR removal is satisfied. By trial and error, we set $T_{r1} = 0.85$ and $T_{r2} = 1$, with $T_{r2}$ being larger to ensure only clearly non-fully-emerged ARs are excluded. 

Due to the relatively strong solar activity in cycle 23, the MDI synoptic maps contain many activity complexes, complicating the removal of both non-fully-emerged and decayed repeat ARs. The reliability of previous detections of new ARs from activity complexes during cycle 23 remains uncertain \citep{Wang2024, Jiang2015}. Additionally, calibrating the polar field or dipole strength between MDI and HMI maps is problematic \citep{sun2018}. Therefore, we restrict our simulation to HMI synoptic maps, which cover nearly the entire cycle 24 and the current cycle 25. Only ARs detected in HMI maps are assimilated, spanning CRs 2097-2290 (May 2010 to November 2024).

ARs are assimilated according to their detection time in synoptic maps, corresponding to their central meridian crossing. Since ARs near the map boundaries are more likely to be only partially observed, a stricter flux balance criterion is applied, requiring the ratio ($r$) of unsigned flux between two polarities to satisfy $0.5 < r < 2$. Before assimilation, each AR is adjusted to ensure balanced unsigned flux between the two polarities. Each polarity is scaled by a factor so that both equal the mean of the two polarities original unsigned fluxes \citep{2020Yeates}. The AR maps used in the simulations are also available in our database \citep{Database_WangRH_2025}.

\section{results}
\subsection{Simulation of the Solar Surface Magnetic Field from 2010 to 2024} \label{sec:sml surfaceflux}

\begin{figure}[htbp!]
\centering
\includegraphics[scale=0.33]{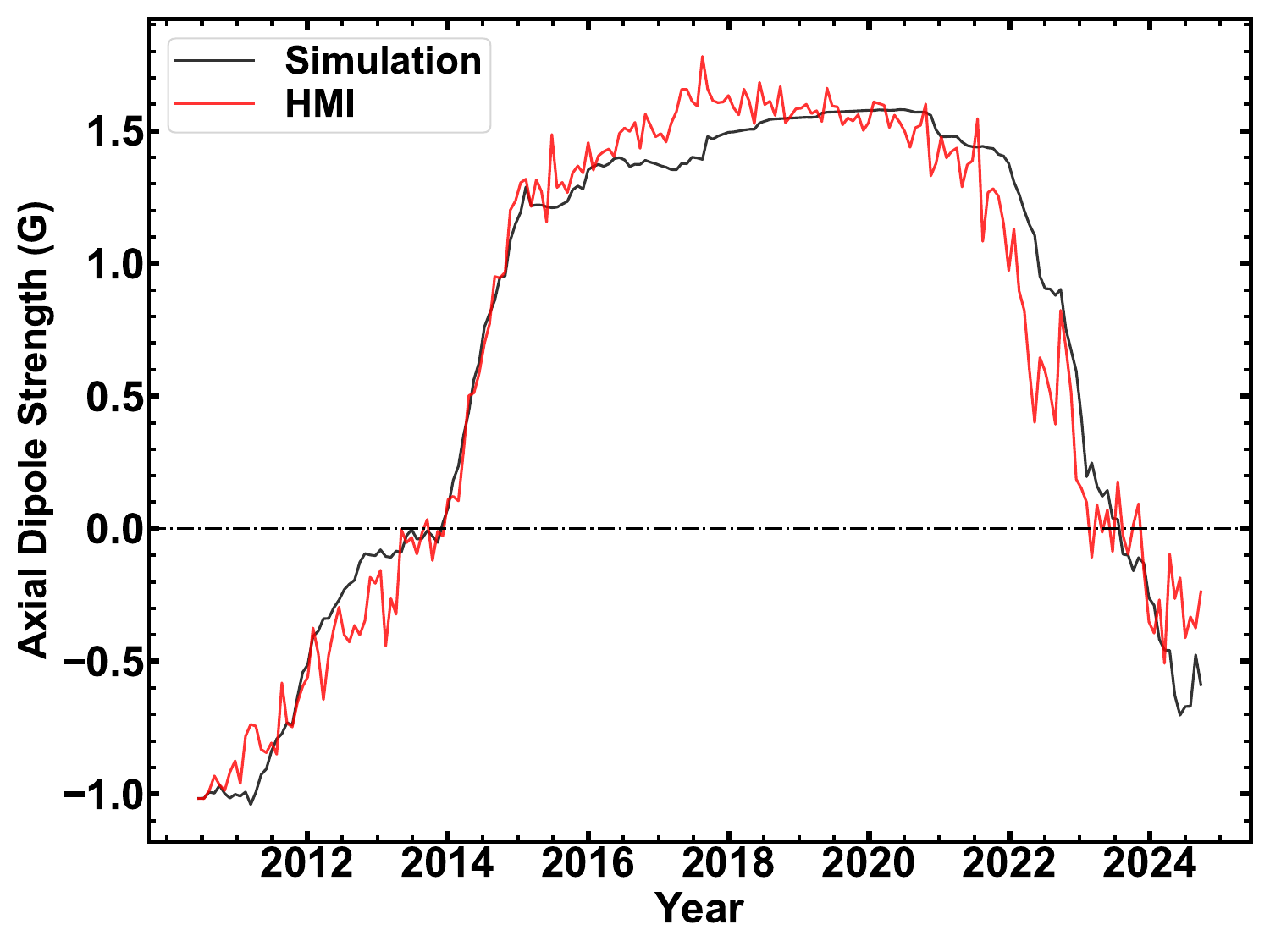}
\caption{Comparison of axial dipole strength obtained from the SFT simulation and from HMI synoptic magnetograms over the period 2010-2024. The horizontal dash-dotted line indicates the zero baseline. 
\label{fig:DpS}}
\end{figure}

\begin{figure*}[htbp!]
\centering
\includegraphics[scale=0.5]{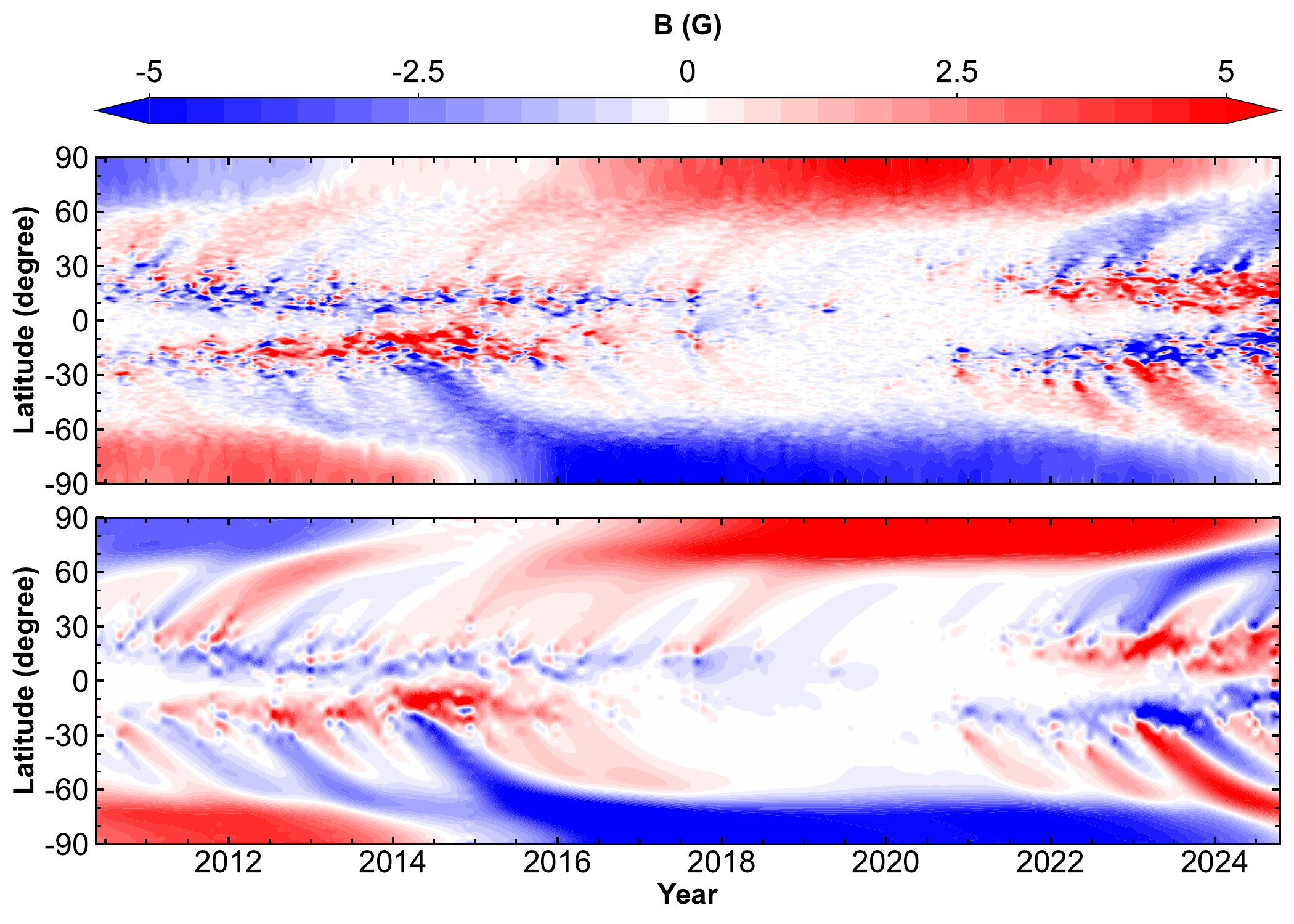}
\caption{Comparison of magnetic butterfly diagrams from 2010 to 2024: HMI synoptic magnetogram observations (top) and SFT simulation results (bottom). HMI data are interpolated to match the simulation spatial resolution.
\label{fig:BFDiagram}}
\end{figure*}

\begin{figure}[htbp!]
\centering
\includegraphics[scale=0.33]{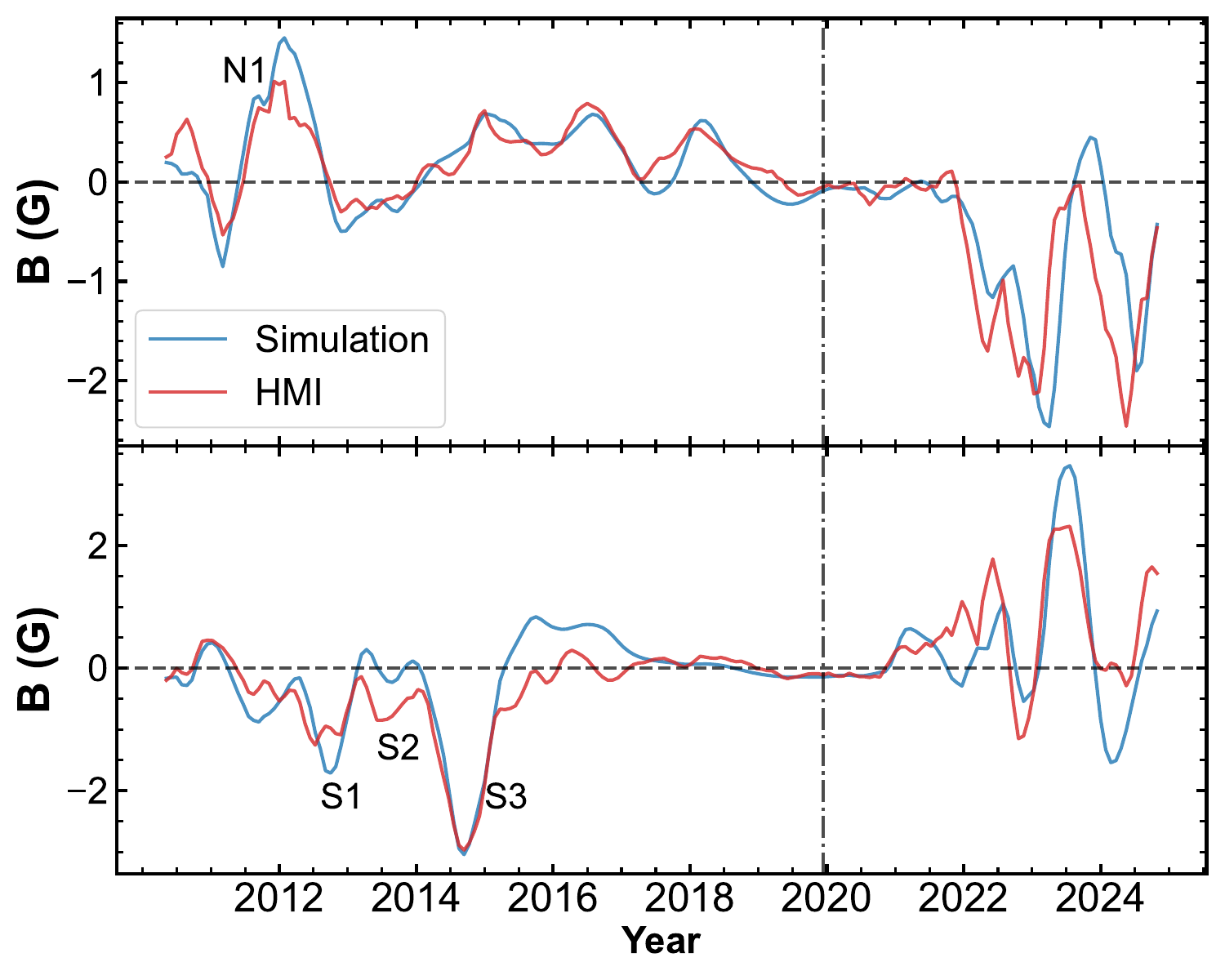}
\caption{Comparison of latitude-averaged magnetic fields ($30^\circ$-$50^\circ$): HMI observations vs. SFT simulations. The top and bottom panels correspond to the northern and southern hemispheres, respectively. The vertical dash-dotted line marks the onset of cycle 25, and the horizontal dashed line indicates the zero baseline. Key surges during cycle 24 are annotated.
\label{fig:meanB}}
\end{figure}

Using the HMI synoptic magnetogram of CR 2097 as the initial condition, we simulate the evolution of the solar surface magnetic field from CR 2097 to CR 2290 (May 2010-November 2024). To validate the simulation, we compare our results with pole-filled HMI synoptic magnetograms. First, we evaluate the solar axial dipole strength, defined as:
\begin{equation}
D=\frac{3}{4\pi} \iint B(\theta ,\phi )\sin\theta \cos\theta d\theta d\phi,
\label{eqDi}
\end{equation}
where $\theta$ and $\phi$ denote colatitude and longitude, respectively, and $B(\theta ,\phi )$ represents the radial component of the surface magnetic field. The axial dipole strength serves as a proxy for the global polar field strength. Figure \ref{fig:DpS} compares the simulated dipole strength with that calculated from HMI observations. 

Overall, our simulation results demonstrate excellent agreement with HMI observations, particularly during 2014-2015, as evidenced by a high correlation coefficient (r=0.98). The HMI-derived dipole strength exhibits fluctuations around zero during polarity reversals in both cycles 24 and 25. For cycle 24, the reversal occurs between around June 2013 and January 2014, consistent with the findings of \cite{Sun2015}, while cycle 25's reversal is observed around June 2023. Our simulation accurately reproduces these reversal timings, though it does not capture the fluctuations around zero seen in HMI observation for cycle 25. 

We further evaluate the simulated butterfly diagram against HMI observations, shown in Figure \ref{fig:BFDiagram}. The simulation shows good overall agreement with HMI data across three key aspects: activity belts, polar field, and surges. 

Since we directly assimilate detected ARs from HMI maps into our simulation, the activity belts closely match observations, as expected. Regarding the polar field above $60^\circ$ latitude in the magnetic butterfly diagram, our simulation reproduces the polarity reversals well, as shown in Figure \ref{fig:BFDiagram}. Both the simulation and observations indicate that the southern polar field of cycle 24 reverses polarity around January 2015, while the polar fields in both hemispheres for cycle 25 approach reversal by November 2024 (the simulation endpoint). However, we observe a discrepancy in the northern polar field reversal timing during cycle 24, with the simulation yielding a reversal around June 2014 compared to approximately June 2013 in HMI maps. Notably, observations from the Wilcox Solar Observatory reveal multiple reversals of the northern polar field between 2012 and 2015, with a final transition to positive polarity around 2015 \citep{WangYM2017}. This suggests that the discrepancy of the northern polar field between our simulation and the HMI data during 2012-2015 may stem from limitations in the HMI polar field measurements.

Our simulation also successfully reproduces the major flux migrations from the activity belts to the poles. Figure \ref{fig:meanB} shows the average magnetic field between $30^\circ$ and $50^\circ$ latitude, where poleward flux migrations primarily occur. Observations indicate that during cycle 24, the dominant polarity of these migrations is positive in the northern hemisphere and negative in the southern hemisphere, with the pattern reversing in cycle 25. Notably, the dominant negative flux migrations in the southern hemisphere during cycle 24 are mainly confined to 2011-2016, followed predominantly by weak positive flux migrations in the declining phase. The simulation accurately captures these observed polarity patterns. It also successfully reproduces two key southern hemispheric surges: the 2014-2016 surge that triggers the reversal of the cycle 24 south polar field, and the 2023-2025 surge that likely drives the reversal of the cycle 25 south polar field. 

While the simulated magnetic butterfly diagram generally agrees with observations, some differences remain. Some small-scale flux is missing in the simulation, particularly during the minimum between cycles 24 and 25. Additionally, the simulated polar fields appear more concentrated than observed, and the surges tend to be stronger. The missing flux likely arises from limitations in the ARISE database, including the low temporal resolution of synoptic maps, the absence of far-side observations, and challenges in identifying newly emerged active regions within activity complexes \citep{Wang2024}. The differences in polar field distribution are likely related to the meridional flow profile adopted in our simulation and may be improved by employing a more suitable profile. The cause of the stronger simulated surges remains uncertain and requires further investigation; possible factors include the choice of transport parameters and the diffusion approximation of supergranulation used in the SFT model.

Several studies have simulated the solar surface magnetic field by assimilating ARs from HMI maps, such as \cite{2020Yeates}, \cite{WangZF2020}, and \cite{YangSH2024}. These works successfully reproduce the magnetic butterfly diagram and the evolution of dipole strength or polar fields, but their simulations are limited to cycle 24. \cite{Dash2024} extends this approach to cover 2010-2023, encompassing both cycle 24 and the ongoing cycle 25. While their simulated butterfly diagram aligns reasonably well with observations, their dipole strength is significantly overestimated during cycle 25. This discrepancy likely arises from their use of idealized BMRs to represent observed ARs. In contrast, our simulation for 2010-2024 achieves better agreement with observations by directly assimilating accurately detected ARs from our ARISE database. Additionally, our good simulation of the surface magnetic field during cycle 24 and the ongoing cycle 25, achieved without radial diffusion or cycle-dependent variation in meridional flow peak speed, indicates that these processes have limited influence on multi-cycle surface field evolution.

\subsection{Investigating the Anomalous Southern Poleward Flux Transport in cycle 24 }\label{sec: c24 south}

While flux migrations from the activity belts to the poles are typically dominated by the following polarities (as seen in the northern hemisphere during cycle 24 and both hemispheres during cycle 25), Figures \ref{fig:BFDiagram} and \ref{fig:meanB} reveal a distinct pattern in the southern hemisphere during cycle 24. Here, following-polarity (negative) migrations primarily occur between 2011 and 2016, with the subsequent declining phase (2016-2020) dominated by weak leading-polarity (positive) flux migrations. This section researches the AR parameters of cycle 24 to investigate the underlying mechanisms driving this anomalous flux migration pattern.

\subsubsection{Active region Characteristics in the Southern Hemisphere during Cycle 24} \label{sus:DiDf}

\begin{figure}[htbp!]
\centering
\includegraphics[scale=0.37]{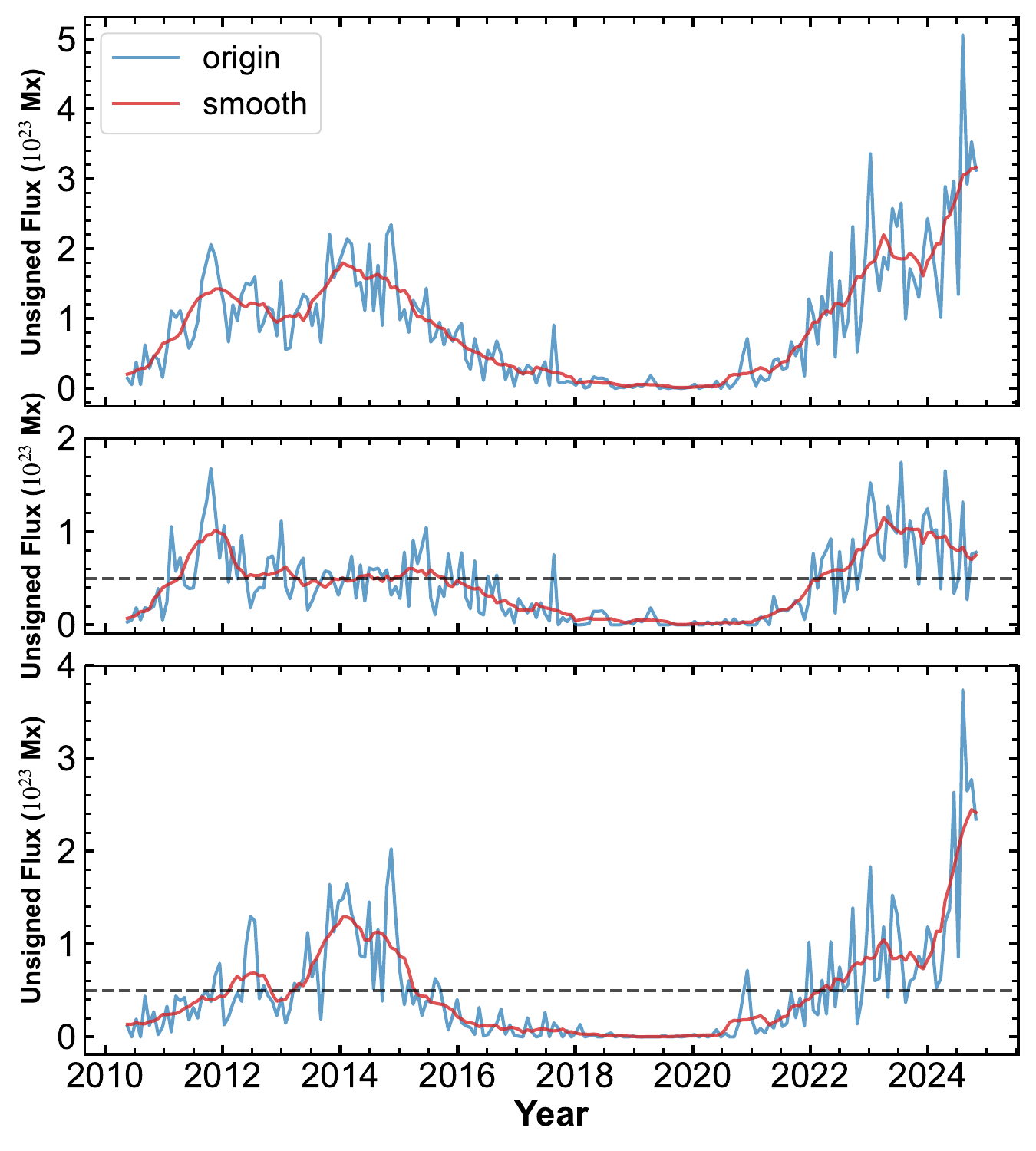}
\caption{Temporal evolution of unsigned AR flux (2010-2024). From top: total flux, northern hemisphere flux, and southern hemisphere flux. Red lines indicate 9-CR smoothed values. Dashed lines indicate the typical value ($0.5 \times 10^{23}$ Mx) of AR flux in the northern hemisphere.
\label{fig:usflux}}
\end{figure}

\begin{figure*}[htbp!]
\centering
\includegraphics[scale=0.5]{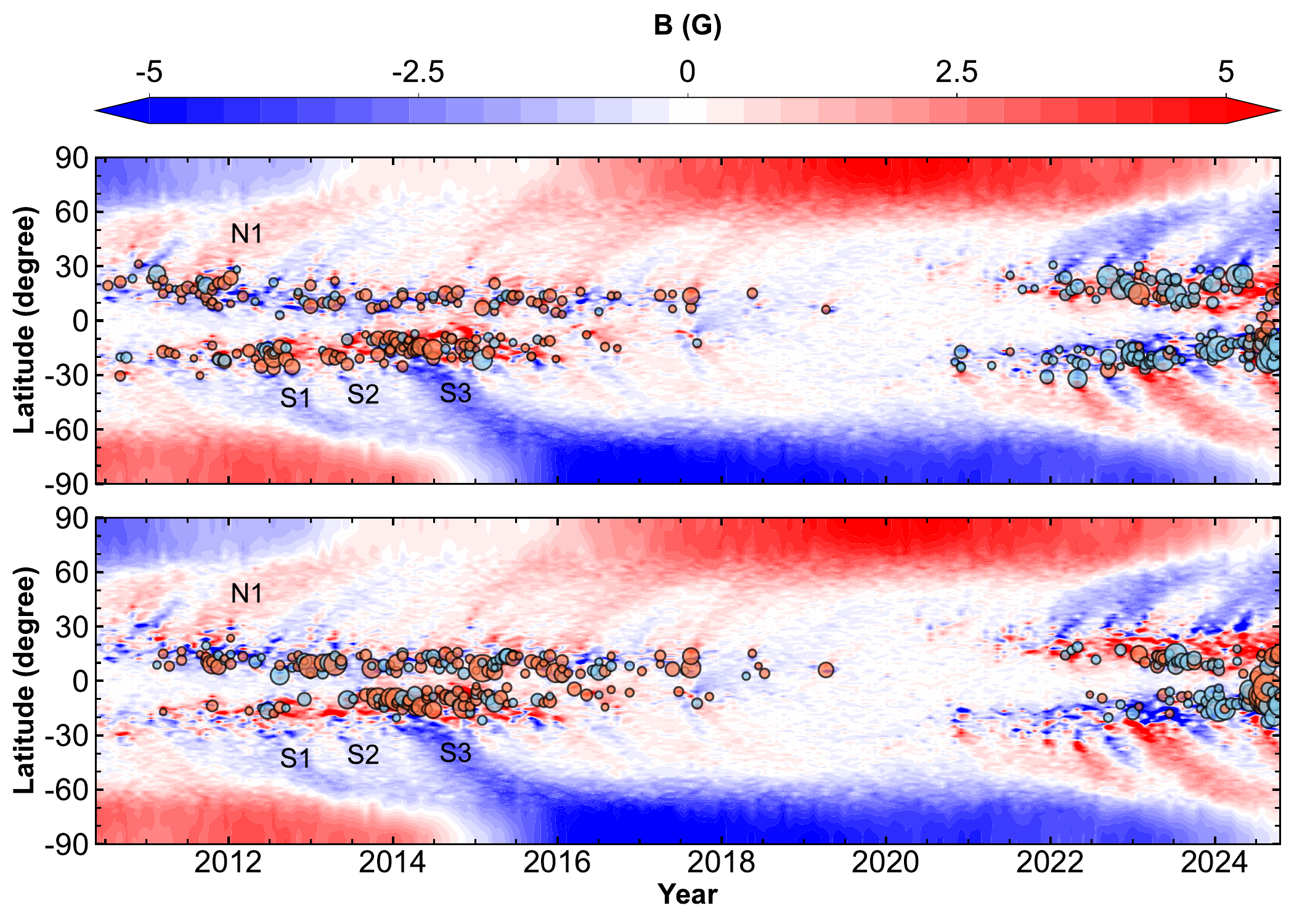}
\caption{Magnetic butterfly diagrams overplotted with AR initial axial dipole strength (top) and final axial dipole strength (bottom). Only ARs with axial dipole strength exceeding 0.01 G are displayed. Circle size is proportional to the value, and color indicates polarity (orange for positive, blue for negative).
\label{fig:BD&DF}}
\end{figure*}

To investigate the origin of the anomalous southern poleward flux migrations during cycle 24, we analyze AR parameters from the ARISE database \citep{Database_WangRH_2025}. Figure \ref{fig:usflux} presents the unsigned magnetic flux of southern hemisphere ARs, comparing them with the northern hemisphere and total unsigned fluxes. Our flux measurements show good agreement with both sunspot area data from \cite{Mordvinov2016} and AR flux measurements in \cite{zhukova2024}, which employed the Crimean Astrophysical Observatory's magneto-morphological classification catalog \citep{Abramenko2021}.

The strength of surges is influenced by AR properties such as latitudinal polarity separation and total flux, as well as by surface transport parameters including meridional flow profile, flow speed, and supergranular diffusivity \citep{WangYM1989ApJ, WangYM2009, Sun2015}. Our previous work \citep{Wang2024} has demonstrated that the BMR approximation gives nearly the same initial axial dipole strength ($D_i$) as the original ARs. When approximating ARs as BMRs, their $D_i$ is given by 
\begin{equation}
D_i = \frac{3}{4 \pi R_\odot^2} \Phi d_\lambda \cos \lambda,
\end{equation}
where $\lambda$ is latitude, $\Phi$ is the unsigned flux of one polarity, and $d_\lambda$ represents the latitudinal separation between two polarities \citep{Wang1991}. Since variations in surface transport parameters are usually neglected \citep{2020Yeates, Dash2024}, we also treat them as constant in our analysis. Including latitudinal separation and AR flux, the initial axial dipole strength ($D_i$) can serve as a proxy for the impact of AR on the surge. Final axial dipole strength ($D_f$) quantifies the AR's net contribution to the polar field after the surface flux transport. According to \cite{Wang2024}, $D_i$ and $D_f$ for most ARs are smaller than 0.01 G.  Since these ARs usually have negligible influence, we only focus on ARs with either $|D_i|$ or $|D_f|$ exceeding 0.01 G. Figure \ref{fig:BD&DF} displays the axial dipole strengths of those selected ARs from 2010 to 2024 overplotted on the magnetic butterfly diagram.

Figure \ref{fig:BD&DF} reveals a strong correlation between AR initial axial dipole strengths $D_i$ and surges, with clusters of $D_i$ typically appearing at surge onsets. The hemispheric distribution exhibits a pronounced asymmetry. While the northern hemisphere $D_i$ and $D_f$ values maintain relatively uniform temporal distributions, the southern hemisphere exhibits apparent clustering near solar maximum, particularly for $D_f$. These southern hemisphere AR clusters coincide with the AR flux peaks of cycles 24 and 25 shown in Figure \ref{fig:usflux}. The southern flux peaks dominate the global AR flux maximum of two cycles respectively. Notably, the southern hemisphere flux peak in cycle 25 substantially exceeds both the peak in cycle 24 and concurrent northern hemisphere values. Since our primary interest lies in the anomalous southern surges of cycle 24, we will focus our subsequent analysis on cycle 24 AR parameters.

In the northern hemisphere during cycle 24, AR flux exhibits a prominent peak around September 2011, as shown in Figure \ref{fig:usflux}. During this period, many ARs with $|D_i| > 0.01$ G also appear in Figure \ref{fig:BD&DF}. These ARs trigger a significant surge, N1, visible in Figures \ref{fig:meanB} and \ref{fig:BD&DF}, which leads to the reversal of the northern polar field. Since these ARs emerged at relatively high latitudes and $D_f$ tends to decrease with increasing latitude \citep{Jiang2014apj}, they do not exhibit large $D_f$ values. Apart from this peak, the AR flux, $D_i$, and $D_f$ remain relatively uniform, with AR flux maintaining a nearly constant level of approximately $0.5 \times 10^{23}$ Mx, except during the rising and declining phases.

In contrast to the northern hemisphere, the southern hemisphere during cycle 24 exhibits significantly intermittent and nonuniform distributions of AR $D_i$, $D_f$, and flux, as seen in Figures \ref{fig:usflux} and \ref{fig:BD&DF}. The AR flux displays a major peak during 2014-2015 and two secondary peaks around June 2012 and June 2013. Similarly, the AR $D_i$ also displays a larger cluster during 2014-2015 and two smaller clusters around June 2012 and June 2013. The ARs emerging during 2014-2015 generate the prominent surge S3 (2014-2016), while those from the secondary peaks produce weaker negative surges S1 and S2. These three surges represent the primary following-polarity (negative) surges in the southern hemisphere. Due to their relatively high emergence latitudes in June 2012 and June 2013, the ARs during these periods have small $D_f$ values and thus small impacts on the polar field. Outside these peaks, the southern AR flux typically remains below $0.5 \times 10^{23}$ Mx, the characteristic flux value for the northern hemisphere. 

Despite the notable differences in flux emergence patterns between the two hemispheres, their total unsigned AR fluxes in cycle 24 are nearly identical: $47.75 \times 10^{23}$ Mx for the north and $48.19 \times 10^{23}$ Mx for the south.

\subsubsection{Analysis of Southern Hemisphere Active Regions in 2014-2015}\label{sus:smlt AR evolution}

\begin{figure}[htbp!]
\centering
\includegraphics[scale=0.33]{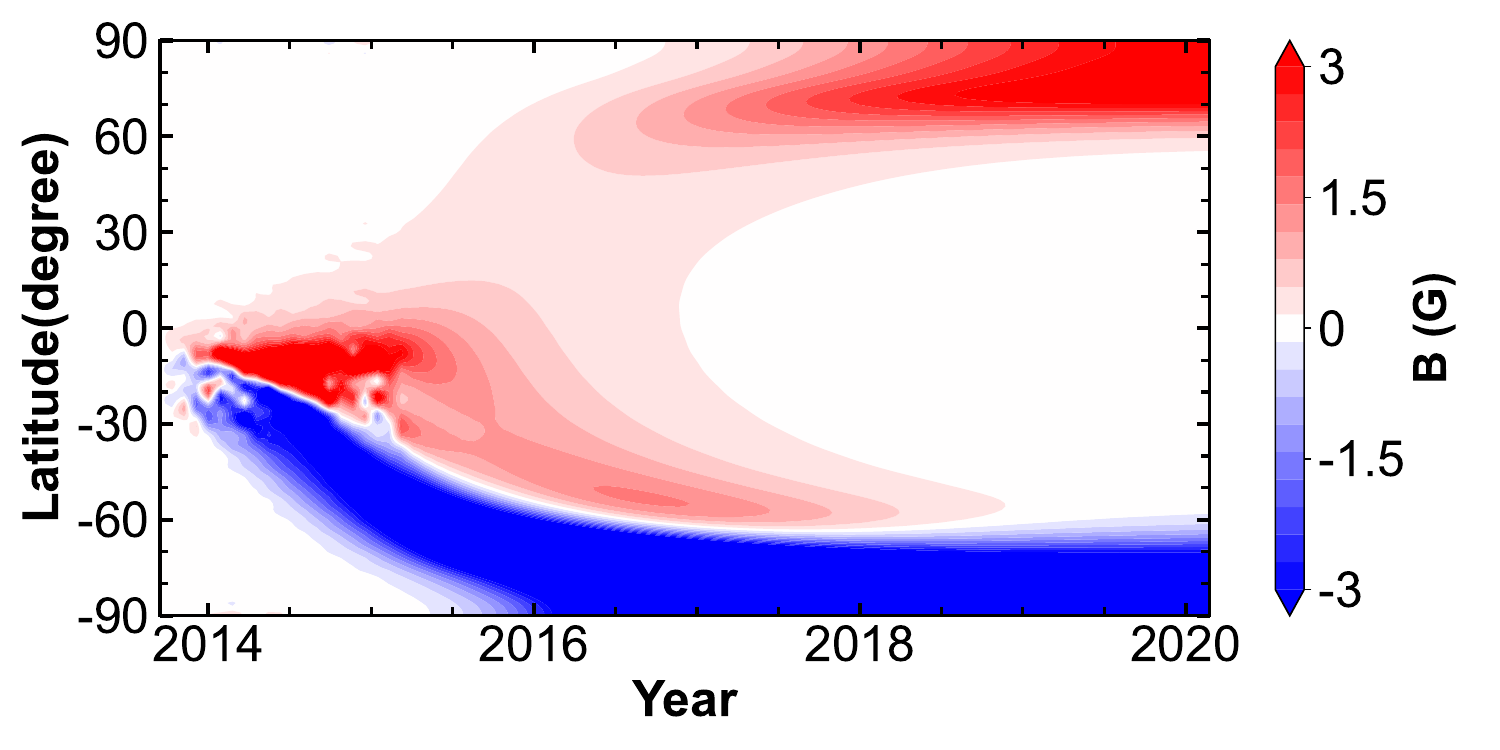}
\caption{Time-latitude evolution of simulated magnetic fields from ARs that emerged in the southern hemisphere during CRs 2141-2160. ARs are sourced from the ARISE database \citep{Database_WangRH_2025} and detected from the HMI synoptic magnetograms.
\label{fig:BF14-20}}
\end{figure}

Our analysis reveals that the AR initial axial dipole strength ($D_i$), final axial dipole strength ($D_f$), and AR unsigned flux in the southern hemisphere during cycle 24 are strongly concentrated during 2014-2015. Based on their distributions presented in Figures \ref{fig:usflux} and \ref{fig:BD&DF}, we identify the primary AR emergence period as CRs 2141-2160, spanning from September 2013 to February 2015. These ARs cause the prominent southward surge (S3) during 2014-2016, as shown in Figure \ref{fig:BD&DF}, and are hereafter referred to as S3-ARs. The combined unsigned magnetic flux of the S3-ARs is $22.15 \times 10^{23}$ Mx, accounting for 46\% of the total unsigned flux from all southern hemisphere ARs in cycle 24. To further research the S3-ARs, we apply the SFT model described in Section \ref{sec:model} to simulate their evolution. Our simulation initializes with zero magnetic fields and assimilates these ARs at their respective central meridian passage times.

Figure \ref{fig:BF14-20} illustrates the simulated magnetic field evolution of the S3-ARs. It demonstrates that while their following polarities drive the prominent southward surge S3, a portion of the leading polarity flux also migrates poleward. Although this leading-polarity flux is eventually canceled by the following-polarity flux, it still contributes to the positive flux migration observed in the southern hemisphere during 2015-2017, as shown in Figure \ref{fig:BD&DF}. 

Beyond the flux transported to the south pole, a substantial portion of the leading-polarity flux from the S3-ARs crosses the equator and ultimately reaches the north pole, as shown in Figure \ref{fig:BF14-20}. An equivalent amount of following-polarity flux ultimately remains at the south pole. Consequently, these ARs significantly influence the polar field evolution. Our simulation indicates that they contribute approximately 1 G to the solar axial dipole strength. For comparison, the total axial dipole strength generated by all ARs emerging in the southern hemisphere during cycle 24 is about 1.05 G. Thus, the S3-ARs alone account for nearly the entire net dipole contribution from all southern ARs in cycle 24. 

The origin and characteristics of the southward surge S3 were previously investigated by \cite{WangZF2020}. Their analysis identified ARs emerging during CRs 2145-2159 as the dominant sources of the surge, with two distinct activity complexes contributing most of its strength. These results support our findings and indicate a non-uniform longitudinal distribution of AR emergence during 2014-2015. In the following, we conduct a detailed investigation of the longitudinal emergence characteristics of the S3-ARs.

\begin{figure}[htbp!]
\centering
\includegraphics[scale=0.36]{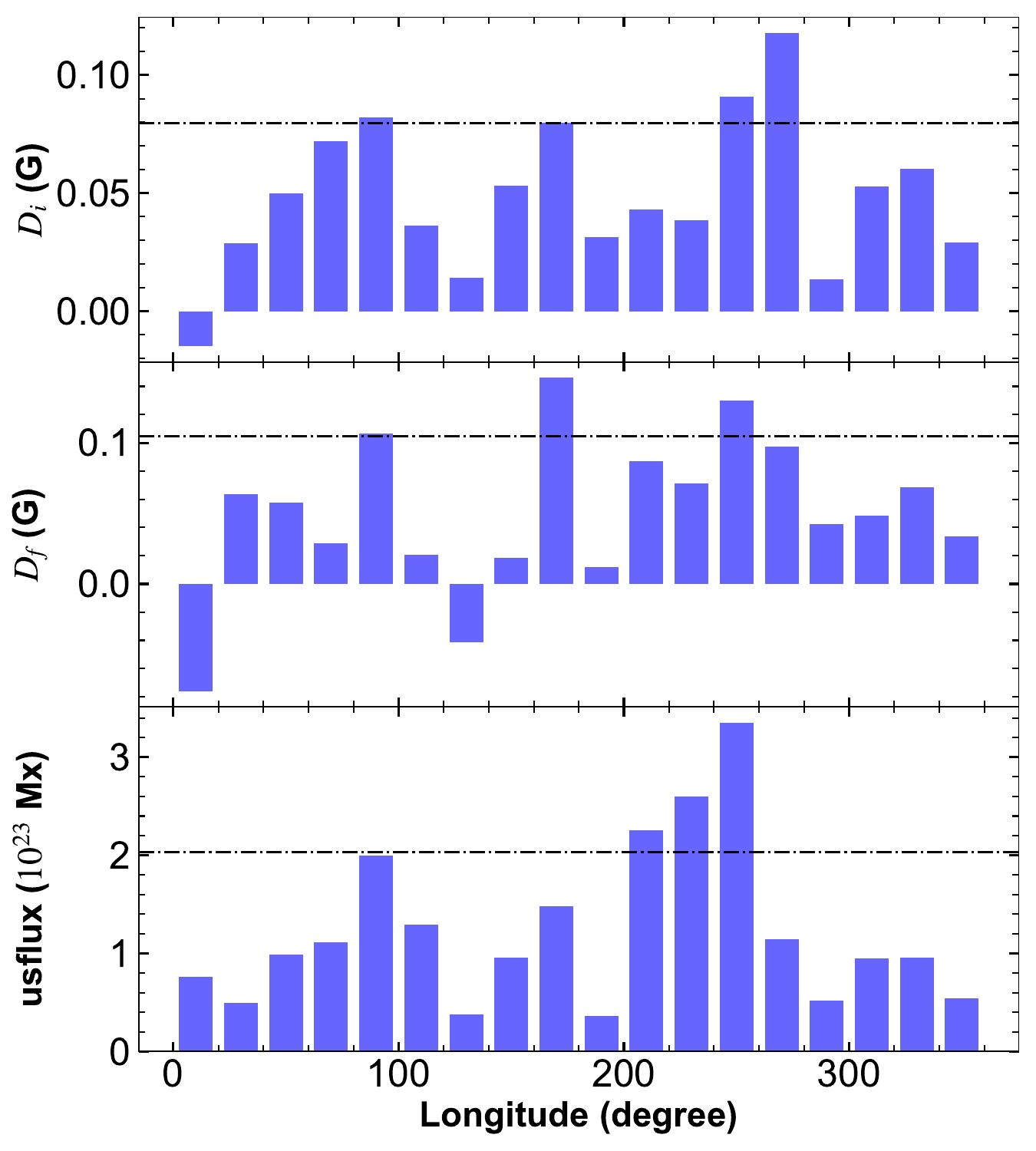}
\caption{Longitude distribution of the S3-ARs parameters: initial axial dipole strength ($D_i$, top), final axial dipole strength ($D_f$, middle), and unsigned flux (bottom). The dash-dotted line in each panel marks the critical threshold defined as the mean value plus one standard deviation across all longitude bins.
\label{fig:lonDisbt}}
\end{figure}

\begin{table}[htbp]
\caption{AR parameters in the active longitudes}
\label{tab:AL}
\begin{threeparttable}
\begin{tabular}{ccccc}
\hline \hline
lon     & num & unsigned flux  & $D_i$ & $D_f$ \\ 
 (degree)  &  & ($10^{23}$ Mx) & (G)   & (G)   \\ \hline
80-100  & 9   & 2.00           & 0.082   & 0.107        \\
200-260 & 36  & 8.21           & 0.173   & 0.288         \\
all     & 159 & 22.15          & 0.879   & 0.915         \\ \hline
\end{tabular}

\end{threeparttable}
\end{table}

Figure \ref{fig:lonDisbt} shows the longitudinal distributions of initial axial dipole strength ($D_i$), final axial dipole strength ($D_f$), and unsigned flux for the S3-ARs, analyzed in $20^\circ$ bins to reduce fluctuations. The unsigned flux distribution reveals two distinct active longitude bands: a primary concentration between $200^\circ$ and $260^\circ$, where flux values significantly exceed the threshold (defined as the mean plus one standard deviation), and a secondary peak between $80^\circ$ and $100^\circ$. While the unsigned flux in the $80^\circ$-$100^\circ$ range is close to the threshold, it remains notably higher than in other longitude bins. Axial dipole strengths (both $D_i$ and $D_f$) in these two active longitude ranges also show enhancement, though less pronounced than the unsigned flux. A notable increase in axial dipole strength is also observed in the $160^\circ$-$180^\circ$ range. Additionally, ARs located between $0^\circ$ and $20^\circ$ exhibit significantly negative $D_f$ values. The weak correlation between flux and axial dipole strength distributions is likely due to the influence of multiple factors beyond flux alone, such as AR latitude, latitudinal polarity separation, and other parameters, which together determine the axial dipole strength \citep{Wang2024}.

The values of AR parameters within the two active longitudes as well as across all longitudes are summarized in Table \ref{tab:AL}. Although only 45 ARs, representing 28\% of the total, are located within these two active longitudes, they account for nearly half of the total unsigned flux, 29\% of the net $D_i$, and 43\% of the net $D_f$. The concentration of $D_i$ within the active longitudes is weaker compared to that of the unsigned flux and $D_f$. \cite{Berdyugina2003} studied active longitudes based on sunspot area and found two active longitudes separated by $180^\circ$. The two active longitudes identified in our analysis are also approximately $180^\circ$ apart, consistent with their findings.

\subsubsection{A Possible Reason for the Positive Flux Migrations in the Southern Hemisphere after 2016}\label{sus:posMig}

\begin{figure}[htbp]
\centering
\includegraphics[scale=0.33]{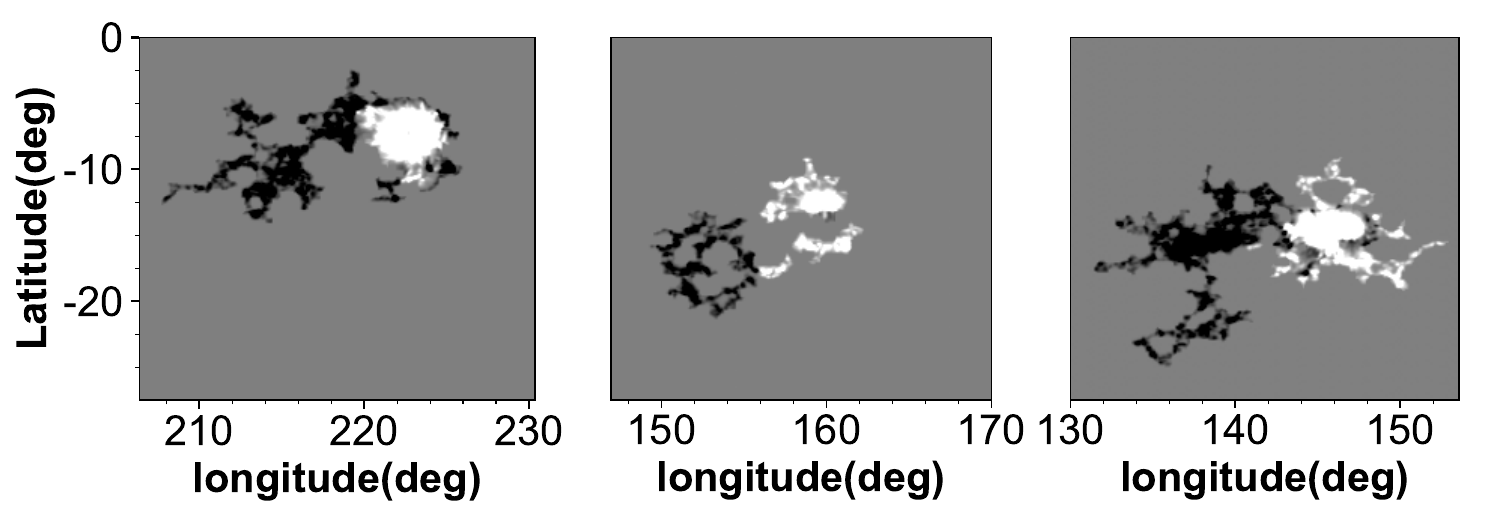}
\caption{The three major ARs in the southern hemisphere from 2016 to 2020 with initial axial dipole strength $D_i > 0.01$ G. From left to right: AR 2177-3 (NOAA 12546), AR 2180-2 (NOAA 12576), and AR 2182-1 (NOAA 12599), where the numeric labels (e.g., 2177-3) refer to identifiers in our AR database.
\label{fig:ARs1617}}
\end{figure}


Flux migrations are typically dominated by the following polarity, as observed in the northern hemisphere during cycle 24 and in both hemispheres during cycle 25. In contrast, in the southern hemisphere of cycle 24, following-polarity (negative) migrations are mainly limited to 2011-2016. After 2016, both observed and simulated butterfly diagrams show that flux migrations are primarily of the leading polarity (positive).  The leading-polarity migrations are usually linked to anti-Joy or anti-Hale ARs, which have negative axial dipole strengths in cycle 24. However, ARs emerging between 2016 and 2020 predominantly exhibit positive initial and final axial dipole strengths. As shown in Figure \ref{fig:BD&DF}, most ARs with $|D_i| > 0.01$ G or $|D_f| > 0.01$ G are positive. The three main ARs with $D_i > 0.01$ G in southern hemisphere ARs during 2016-2020 are shown in Figure \ref{fig:ARs1617}. The total axial dipole strengths of all southern ARs during 2016-2020 are also positive: about 0.04 G for $D_i$ and about 0.09 G for $D_f$. These results suggest that the positive flux migrations after 2016 are not driven by anti-Joy or anti-Hale ARs.

We have shown that the positive poleward migrations during 2015-2017 are caused by the leading-polarity flux from S3-ARs. For the positive migrations observed between 2017 and 2020, we suggest they are likely driven by the poleward transport of leading-polarity flux from regular ARs. This interpretation is supported by the broader latitudinal distribution of leading-polarity migrations compared to the following polarity in the southern hemisphere, as illustrated in Figure \ref{fig:BF14-20}. During AR evolution, the following polarity typically locates at higher latitudes due to the tilt angle, while the leading polarity tends to distribute across a wider latitude range than the following polarity at lower latitudes. Similar flux transport behavior has been demonstrated in previous simulations of individual ARs \citep{Yeates2015, WangYM2017, Jiang2019}.

Since ARs emerging after 2016 are relatively weak, as seen in Figure \ref{fig:usflux}, the flux migrations of both polarities are weaker compared to those of ARs that emerged between 2011 and 2016. As a result, the broader positive migration from the leading polarity flux becomes more prominent in the full magnetic butterfly diagram. Additionally, Figure \ref{fig:BD&DF} shows narrow negative migrations that precede the broader positive ones, which also support our analysis.

ARs typically emerge continuously and generally follow Joy’s law and Hale’s law, as observed in the northern hemisphere during cycle 24. In such cases, the poleward-migrating leading-polarity flux from earlier ARs is often canceled by the following-polarity flux of subsequently emerging ARs. As a result, the magnetic butterfly diagram primarily displays flux migrations of the following polarity. However, after 2016 in the southern hemisphere of cycle 24, AR emergence becomes infrequent, with relatively long intervals between successive emergence. Consequently, the leading-polarity flux has sufficient time to migrate poleward and spread across a broad latitude range before cancellation by the following-polarity flux of later ARs. Since the leading-polarity flux is distributed over a broader latitude range than the following polarity, it becomes more prominent in the butterfly diagram.

\section{Conclusion and Discussion} \label{sec:conclusion}

In this paper, we perform a continuous simulation of the solar surface magnetic field from 2010 to 2024 using the SFT model, with assimilated ARs from our ARISE database \citep{Database_WangRH_2025} as the source. We then analyze the anomalous poleward flux migrations observed in the southern hemisphere during cycle 24.

Some ARs with strong magnetic flux can persist for multiple CRs, and their repeated observations can significantly influence the SFT simulation. In previous work, we implemented a repeat-AR-removal module to eliminate the repeat observation during the decaying phase of ARs. In this study, we apply a similar approach to remove ARs that were still in the process of emerging when first observed. The updated ARISE database now includes 3005 ARs spanning CR 1909 to CR 2290 (1996.5-2024.11). Both the database and associated codes are publicly available on GitHub and Zenodo \citep{Database_WangRH_2025}.

To avoid calibration issues between HMI and MDI magnetograms, we use only HMI data in this study. Assimilating properly detected ARs in our database as the source, we successfully simulate the solar surface magnetic field from 2010 to 2024. The simulation closely matches observations in terms of axial dipole strength and the magnetic butterfly diagram. It also reproduces the polar field reversals timing and poleward surges. Our simulation does not include a radial diffusion term or cyclic variations in flux transport parameters. Notably, \cite{Yeates2025} also achieves a successful reproduction of polar field evolution consistent with proxy observations from 1923 to 1985 without including these terms. These results suggest that the radial diffusion term and variations in flux transport parameters may play a limited role in multi-cycle surface field evolution. The strong agreement between our simulation and observations further demonstrates the reliability of the ARISE database as a source for studying solar surface magnetic field evolution.

Despite the overall good agreement, our simulation still shows some discrepancies compared to observations. Some small flux migrations in the magnetic butterfly diagram are missing, likely due to the limitations in the ARISE database. These limitations stem from the low temporal resolution of synoptic magnetograms, the absence of far-side ARs, and challenges in detecting newly emerged flux within activity complexes. Incorporating far-side ARs has been shown to improve the consistency between SFT simulations and observations \citep{Upton2024, YangD2024}. In addition, the simulated polar fields appear more concentrated than observed, and the simulated surges are stronger. The influence of transport parameters, particularly the meridional flow profile, on these discrepancies will be investigated in future work to improve the simulation.

Both our simulation and observations show that the typically dominant flux migrations of the following polarity, as seen in the northern hemisphere during cycle 24 and in both hemispheres during cycle 25, occur mainly between 2011 and 2016 in the southern hemisphere of cycle 24. After 2016, weak migrations of the leading polarity (positive) become dominant in the southern hemisphere of cycle 24. This anomalous flux transport pattern is attributed to the intermittent emergence of ARs in the southern hemisphere during cycle 24. Most ARs in this hemisphere emerged between 2014 and 2015. In contrast, AR emergence in the northern hemisphere occurred more uniformly except during the rising and declining phases.

For ARs in the southern hemisphere, those emerging during CRs 2141-2160 (September 2013 – February 2015) contribute 46\% of the total unsigned flux and generate a net axial dipole strength comparable to that from all southern ARs in cycle 24. These ARs drive a prominent negative flux surge during 2014-2016, which ultimately reverses the southern polar field. Active longitudes are identified among these ARs, with a strong concentration at $200^\circ$-$260^\circ$ and a secondary peak at $80^\circ$-$100^\circ$. After 2016, AR emergence becomes infrequent, and most ARs follows Joy’s and Hale’s laws. The observed leading-polarity flux migrations after 2016 are not caused by anti-Joy or anti-Hale ARs. Instead, they are likely the result of poleward transport of leading-polarity flux from normal ARs. Due to the long time intervals between successive AR emergence, the leading-polarity flux has sufficient time to spread across a broad latitude range before cancellation by the following-polarity flux of subsequent ARs. This makes the leading-polarity flux stand out more prominently in the butterfly diagram.

The anomalous flux transport pattern in the southern hemisphere during cycle 24 highlights the importance of the temporal distribution of AR emergence in driving poleward flux migrations on the solar surface. \cite{WangZF2022} demonstrated the widespread nonuniformity of poleward flux migrations, which reflects the generally uneven temporal distribution of AR emergence. Although we have shown that concentrated AR emergence can lead to significant surges and that long intervals between AR emergence may result in pronounced surges of leading polarity, the impact of nonuniform AR emergence on the solar surface magnetic field remains to be further investigated.

\cite{2020Yeates}, \cite{Dash2024}, \cite{YangSH2024}, and our simulation all reproduce axial dipole strength and magnetic butterfly diagrams consistent with observations. However, the source data and flux transport parameters used in these studies differ, suggesting that SFT simulations with different sources and transport parameters can still yield similar large-scale features. This indicates that axial dipole strength and visual agreement with butterfly diagrams alone may not be sufficient to independently constrain the source or the transport parameters. These findings highlight the need for additional constraints, such as quantitative comparisons of surface magnetic fields or magnetic power spectra (Luo et al. 2025, in preparation). Moreover, we suggest that the variation in flux transport parameters across previous studies is largely due to the differences in the source data used \citep{Virtanen2017, Whitbread2018}. In addition, transport parameters derived through optimization between SFT simulations and observations, as in \cite{Lemerle2015} and \cite{Whitbread2017}, are also influenced by the choice of source data in the simulation.

The source plays a crucial role in SFT simulations. In future work, we will investigate the impact of different source assimilation methods and AR flux balance approaches. We also plan to examine how various sources influence the optimization of flux transport parameters. Given that meridional flow profiles are poorly constrained by observations, the effects of different profiles on simulations also warrant further study. These studies will enhance our understanding of surface magnetic field evolution and contribute to improving solar cycle prediction. 



\begin{acknowledgements}
We thank the reviewer for the valuable comments and suggestions on improving the paper. The research is supported by the National Natural Science Foundation of China (grant Nos. 12425305, 12173005, and 12350004), National Key R\&D Program of China (grant No. 2022YFF0503800), and SCOSTEP/PRESTO. The SDO/HMI data are courtesy of NASA and the SDO/HMI team. SOHO is a project of international cooperation between ESA and NASA. 
\end{acknowledgements}

\bibliography{sample7}{}
\bibliographystyle{aasjournal}

\end{CJK*}
\end{document}